\documentclass[referee]{cjaa}           % referee version: for submission
\input{epsf.sty}                        %one can use epsf.sty, psfig.sty or both
\input{psfig.sty}

\begin{document}
\title{Dynamical Evolution of Gamma-Ray Burst Remnants with Evolving
Radiative Efficiency}
\author{J. B. Feng\inst{1}, Y. F. Huang\inst{1,2}, Z. G. Dai\inst{1} and T. Lu\inst{1,2}}
\institute{Department of Astronomy, Nanjing University, Nanjing 210093, China
    \and
    LCRHEA, Institute for High-Energy Physics, Chinese Academy of Sciences, Beijing 100039, China\\
    \email{hyf@nju.edu.cn; tlu@nju.edu.cn} }
\date{Received~~2002~~~~~~~~~~~~~~~ ; accepted~~2002~~~~~~~~~~~~~~}

\abstract{ In previous works, a generic dynamical model has been
suggested by Huang et al., which is shown to be correct for both
adiabatic and radiative blastwaves, and in both ultra-relativistic
and non-relativistic phases. In deriving their equations,
Huang et al. have assumed that the radiative efficiency of
the fireball is constant. They then applied their model directly
to realistic cases where the radiative efficiency evolves with
time. In this paper, we abandon the above assumption and
re-derive a more accurate dynamical equation for gamma-ray
burst remnants. Numerical results show that Huang et al.'s  model
is accurate enough in general cases.
 }

\authorrunning{J. B. Feng, Y. F. Huang, Z. G. Dai and T. Lu}
\titlerunning{Dynamical Evolution of GRB Remnants with Evolving
Radiative Efficiency}

\maketitle

%%%%%%%%%%%%%%%%%%%%%%%%%%%%%%%%%%%%%%%%%%%%%%%%%%%%%%%%%%%%%%%%
\section{Introduction}

Although the progenitors of gamma-ray bursts (GRBs) are still
controversial (Cheng \& Dai 2001; Cheng \& Lu 2001b; Lu et al.
2000a, b), it is generally believed that energetic fireballs
should be involved, where baryons are eventually accelerated to
ultra-relativistic speed (Wu et al. 2001). After the main burst
phase, the thin baryonic shell expands at ultra-relativistic speed
into the surrounding matter, producing afterglows in soft bands
(Cheng, Huang, \& Lu 2001; Mao \& Wang 2001a, b; Gou et al. 2001a,
b; Huang, Yang, \& Lu 2001; Zhang \& M\'esz\'aros 2002). For
good recent reviews on afterglow observations and theories, see
van Paradijs, Kouveliotou, \& Wijers (2000) and Cheng \& Lu (2001a).

The dynamics of the gamma-ray burst remnants is different in two
cases in which the remnant expansion is either adiabatic or highly
radiative (Blandford \& McKee 1976, 1977).
However, the conditions under which
the remnant dynamics may be considered adiabatic or radiative are
far from unambiguous and are crucially dependent on poorly known
questions about postshock energy exchange between protons and
electrons (M\'esz\'aros, Rees, \& Wijers 1998). Furthermore, a
partially radiative regime with decreasing radiative efficiency
may exist in realistic fireballs (Dai, Huang, \& Lu 1999). So,
it is necessary to construct a dynamical model that is able
to describe a realistic fireball, i.e., a fireball with evolving
radiative efficiency.

The dynamics of gamma-ray burst remnants has been studied
extensively (Sari 1997; Cohen, Piran, \& Sari 1998; Panaitescu,
M\'esz\'aros, \& Rees 1998; Wei \& Lu 1998; Chiang \& Dermer 1999;
Rhoads 1999; Panaitescu \& M\'esz\'aros 1999; Kobayashi, Piran, \&
Sari 1999; Huang et al. 1998a, b, c, 1999a, b, 2000a, b, c, 2002; Dermer
\& Humi 2001). Especially, a generic dynamical model was proposed
by Huang, Dai and Lu (1999a, hereafter HDL99), which is shown to
be applicable to both ultra-relativistic and non-relativistic
blastwaves, no matter whether they are adiabatic or highly
radiative.  In their derivation, Huang, Dai and Lu implicitly
assumed that the radiative efficiency of the fireball, $\epsilon$,
is a constant during the deceleration. They then generalized their
model to discuss realistic blastwaves, where $\epsilon$ evolves
with time (Huang et al. 2000a, b). In this work we will
inspect their generalization carefully. We first repeat the
derivation of HDL99, but abandoning the $\epsilon \equiv {\rm const}$
assumption. We then compare our result with that of HDL99
numerically. It is found that Huang et al.'s generic model can be
applied to realistic remnants satisfactorily.

%%%%%%%%%%%%%%%%%%%%%%%%%%%%%%%%%%%%%%%%%%%%%%%%%%%%%%%%%%%%%%%
\section{Dynamics}

We assume that after the initial GRB phase, the total energy left
in the fireball is comparable to the radiation energy emitted in
gamma-rays, i.e., $E_0 \sim 10^{51}$--$10^{52}$\,ergs. Denote the
mass of the contaminating baryons as $M_0$, then the fireball
continues to expand at a Lorentz factor of $\eta=E_0/(M_0c^2)$.
Subsequently, at a radius $R_0$, the expansion of the fireball
starts to be significantly influenced by the swept-up medium and
external shock may form (Rees \& M\'esz\'aros 1992). As usual,
$R_0$ is supposed to be
\begin{equation}
R_0=\left(\frac{3E_0}{4\pi n m_pc^2\eta^2}\right)^{1/3}
\end{equation}
where $n$ is the number density of the interstellar medium,
$m_{\rm p}$ is the mass of a proton.

\subsection{Basic dynamical equations}

In HDL99, a generic dynamical model that is applicable in both
ultra-relativistic and non-relativistic phases of GRB afterglows
has been proposed. The key point of the model is a differential
equation
\begin{equation}\label{generic model}
\frac{d \gamma}{d m} = - \frac{\gamma^2 - 1}
       {M_{\rm 0} + \epsilon m + 2 ( 1 - \epsilon) \gamma m},
\end{equation}
where $\gamma$ is the bulk Lorentz factor of the fireball, $m$ is
the swept-up mass. Equation (\ref{generic model}) is derived as
follows. Global conservation of energy implies that
\begin{equation}\label{conservation}
d[\gamma (M_{\rm 0}c^{2} + mc^{2} + U)] = dmc^{2} + \gamma dU_{\rm
rad}.
\end{equation}
Here $U$ is the co-moving internal energy with rest-mass excluded,
$U_{\rm rad}$ is the internal energy that is radiated from the
fireball. If a fraction $\epsilon$ of swept-up kinetic energy is
instantaneously radiated from the fireball, then $dU_{\rm rad} =
-\epsilon (\gamma -1)dmc^{2}$. The internal energy $U$ in the
fireball changes because of the change of the kinetic energy of
the swept-up matter, due to expansion of the fireball and the
energy loss through radiation. Thus, we assume $U = (1-
\epsilon)U_{\rm ex}$, where $U_{\rm ex}$ is the internal energy
produced in this expansion. It is usually assumed that $dU_{\rm
ex} = (\gamma -1)dmc^{2}$. However, the jump conditions (Blandford
\& Mckee 1976) at the forward shock imply that $U_{\rm ex} =
(\gamma -1)mc^{2}$, so the correct expression for $dU_{\rm ex}$
under thin shell approximation should be $dU_{\rm ex} = d[(\gamma
-1)mc^{2}] = (\gamma -1)dmc^{2} + mc^{2}d\gamma$. Assuming
$\epsilon \simeq const$, then from equation (\ref{conservation})
we can obtain equation (\ref{generic model}).

It is worth noting that in the expression of $dU_{\rm ex} =
(\gamma -1)dmc^{2} + mc^{2}d\gamma$, the term $mc^{2}d\gamma$ is
negative when the fireball is decelerating. This term, in fact,
represents the loss of internal energy due to volume expansion of
the fireball, i.e., the adiabatic loss term ($dU_{adi}$) defined
by Dermer and Humi (2001). This can be clearly seen from equation
(13) of Dermer and Humi (2001). Under thin shell approximation,
their equation can be approximately simplified as $mc^{2}d\gamma$.

In the above derivation, $\epsilon$ is assumed to be constant
during the deceleration. However, in realistic fireballs,
$\epsilon$ is expected to evolve from 1 to 0 owing to the changes
in the relative importance of synchrotron-induced and
expansion-induced loss of energy (Dai, Huang, \& Lu 1999).
Equation (\ref{generic model}) has been simply generalized to the
case that $\epsilon$ evolves with time (Huang et al. 1999b, 2000a,
b, c). However, this might induce some errors. Below, we will abandon
the constant $\epsilon$ assumption and derive the equations that
are strictly applicable for fireballs with evolving radiative
efficiency.

The assumption that $U = (1- \epsilon)U_{\rm ex}$ overestimates
the true internal energy, because at late stages $\epsilon$ is
near 0, but at early stages it is about 1. Instead of using $U =
(1 - \epsilon) U_{\rm ex}$, we use the expression that $dU = (1-
\epsilon)dU_{\rm ex}$. Substituting it into equation
(\ref{conservation}), we obtain another differential equation
depicting the evolution of fireballs
\begin{equation}\label{improved generic model}
\frac{d \gamma}{d m} = - \frac{\gamma^2 - 1}
       {M_{\rm 0} + m + U/c^{2} + (1 - \epsilon) \gamma m},
\end{equation}
with
\begin{equation} \label{dudex}
dU = (1 - \epsilon) dU_{\rm ex}
   = (1 - \epsilon) [ (\gamma-1)dmc^{2} + mc^{2}d\gamma ].
\end{equation}

In the highly radiative case ($\epsilon \simeq 1$, and $U =
0$), equation (\ref{improved generic model}) reduces to the case of
Blandford \& McKee (1976)
\begin{equation}\label{highly radiative}
\frac{d \gamma}{d m} = - \frac{\gamma^2 - 1}
       {M_{\rm 0} + m }.
\end{equation}
While in the fully adiabatic case ($\epsilon \simeq 0$, and $U =
U_{\rm ex} = (\gamma - 1)mc^{2}$), equation (\ref{improved generic
model}) reduces to the adiabatic case of HDL99
\begin{equation}\label{adibatic case}
\frac{d \gamma}{d m} = - \frac{\gamma^2 - 1}
       {M_{\rm 0} + 2 \gamma m }.
\end{equation}
In fact, taking $\epsilon \equiv const$, equation (\ref{improved
generic model}) exactly reduces to the generic model of HDL99.
But if $\epsilon$ evolves with time, we would expect that the
fireball described by equation (\ref{improved generic model}) will
decelerate more rapidly than another fireball described by
equation (\ref{generic model}).

\subsection{Radiative Efficiency}

According to Blandford \& McKee (1976), the electron number
density ($n'$) and energy density ($e'$) of the shocked medium in
the frame co-moving with the fireball can be written as
(also see: Huang et al. 1998b)
\begin{equation}\label{electron mumber density}
n'=\frac{\hat{\gamma}\gamma+1}{\hat{\gamma}-1}n\,,
\end{equation}
\begin{equation}\label{internal energy density}
e'=\frac{\hat{\gamma}\gamma+1}{\hat{\gamma}-1}(\gamma-1)nm_pc^2\,,
\end{equation}
where $\hat{\gamma}$ is the adiabatic index of the shocked medium,
which is generally between 4/3 and 5/3. Equations (\ref{electron
mumber density}) and (\ref{internal energy density}) are
appropriate for both relativistic and non-relativistic blastwaves.
From the definition of $\hat{\gamma}$ (Blandford \& McKee 1976),
Dai, Huang, \& Lu (1999) gave a simple and useful approximate
expression for $\hat{\gamma}$: $\hat{\gamma} \simeq
(4\gamma+1)/(3\gamma)$. It can be seen from this approximation
that $\hat{\gamma} \simeq 4/3$ for an extremely relativistic
blastwave and $\hat{\gamma} \simeq 5/3$ for a non-relativistic
shock.

\begin{figure}
   \begin{center}
   \epsscale{1.0}{1.0}          %% rescaled to {x_size}{y_size}, default is {1}{1}
   \plotone{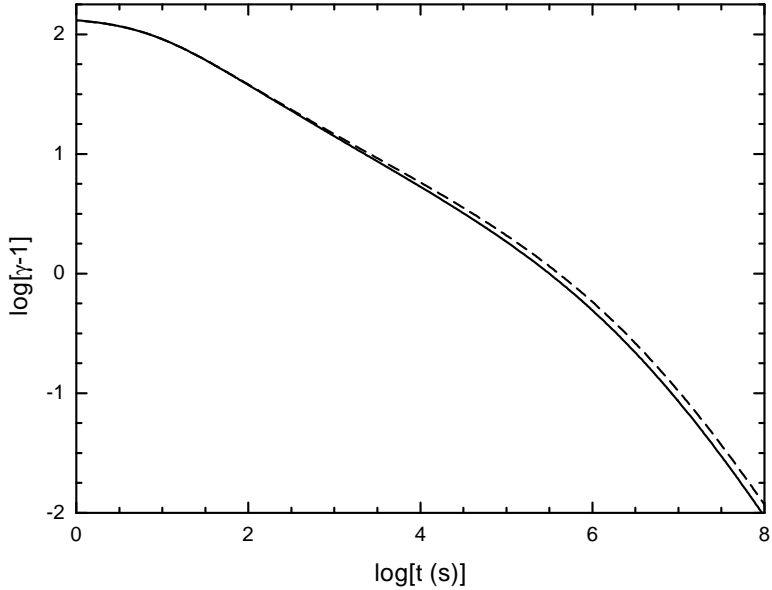}
   \caption{Evolution of the bulk Lorentz factor ($\gamma$). The dashed
   line corresponds to Eq. (\ref{generic model}). The solid line is drawn
   according to Eq. (\ref{improved generic model}).
   Parameters: $E_0 = 10^{52}$ ergs, $n = 1 $ cm$^{-3}$,
   $M_{\rm 0} = 2 \times 10^{-5}M_{\odot}$, $\epsilon_e=1.0$ and $\epsilon_B=0.01$.}
   \label{}
   \end{center}
\end{figure}

As usual, we assume that the magnetic density in the co-moving
frame is a fixed fraction $\epsilon_B$ of the internal energy
density, viz., $ B^\prime = (8 \pi \epsilon_B e^\prime)^{1/2} $,
and that the shock-accelerated electrons behind the blastwave
carry a fraction $\epsilon_e$ of the internal energy
(Huang et al. 2000a, b). This implies
that the minimum Lorentz factor of the random motion of electrons
in the co-moving frame is $\gamma_{e,min} = \epsilon_e(\gamma-1)
m_p/m_e + 1$. We here consider only synchrotron emission from
these electrons, and neglect the contribution of inverse Compton
emission because the latter emission is of minor importance
particularly at late times of the evolution (Waxman 1997; Dai \&
Lu 1998). The energy of a typical accelerated electron behind the
blastwave is lost both through synchrotron radiation and through
expansion of the fireball, thus the radiative efficiency of this
single electron is given by $t^{\prime -1}_{syn}/(t^{\prime
-1}_{syn}+t^{\prime -1}_{ex})$(Dai \& Lu 1998; Dai, Huang, \& Lu
1999), where
$t^\prime_{syn}$ is the synchrotron cooling time, $t^\prime_{syn}
= 6 \pi m_e c / (\sigma_T B^{\prime 2} \gamma_{e,min})$, and
$t^\prime_{ex}$ is the co-moving frame expansion time,
$t^\prime_{ex}=R/(\gamma c)$. Here $R$ is the radius of the
blastwave. Since all of the accelerated electrons behind the
blastwave carry only a fraction $\epsilon_e$ of the internal
energy, the radiative efficiency of the fireball can be given by
(Dai, Huang, \& Lu 1999)
\begin{equation}\label{epsilon}
\epsilon = \epsilon_e \frac {t^{\prime -1}_{syn}} {t^{\prime
-1}_{syn} + t^{\prime -1}_{ex}}.
\end{equation}
In the highly radiative case, $\epsilon_e \simeq 1 $ and $
t^{\prime}_{syn} \ll t^\prime_{ex}$, we have $\epsilon \simeq 1$.
The early evolution of the remnants is likely in this regime. For
an adiabatic expansion, $\epsilon_e \ll 1 $ or $ t^{\prime}_{syn}
\gg t^\prime_{ex}$, we get $\epsilon \simeq 0$. The late evolution
is believed to be in this regime. In realistic case, the radiative
efficiency of the fireball ($\epsilon$) evolves from about 1 to 0
(Huang et al. 2000a).

\begin{figure}
   \begin{center}
   \epsscale{1.0}{1.0}          %% rescaled to {x_size}{y_size}, default is {1}{1}
   \plotone{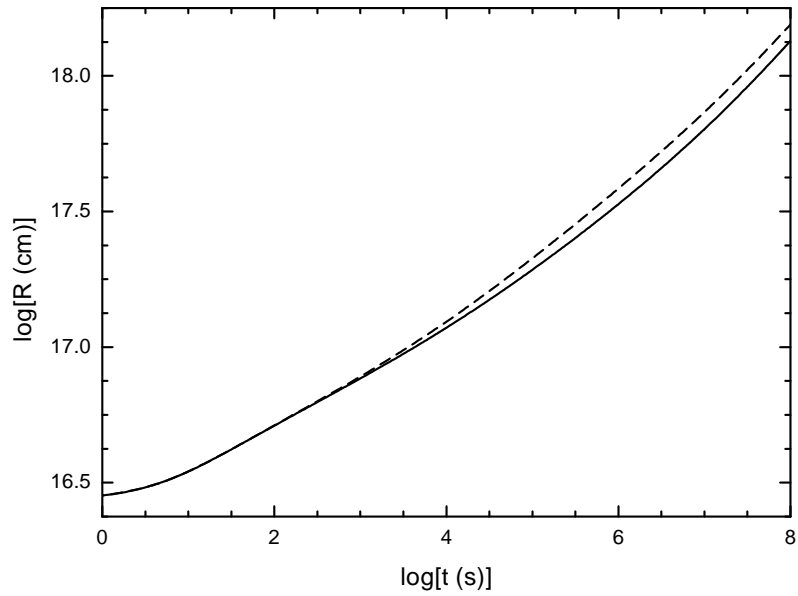}
   \caption{ Evolution of the shock radius ($R$). Parameters and line styles
   are the same as in Fig.~1. }
   \label{}
   \end{center}
\end{figure}

\begin{figure}
   \begin{center}
   \epsscale{1.0}{1.0}          %% rescaled to {x_size}{y_size}, default is {1}{1}
   \plotone{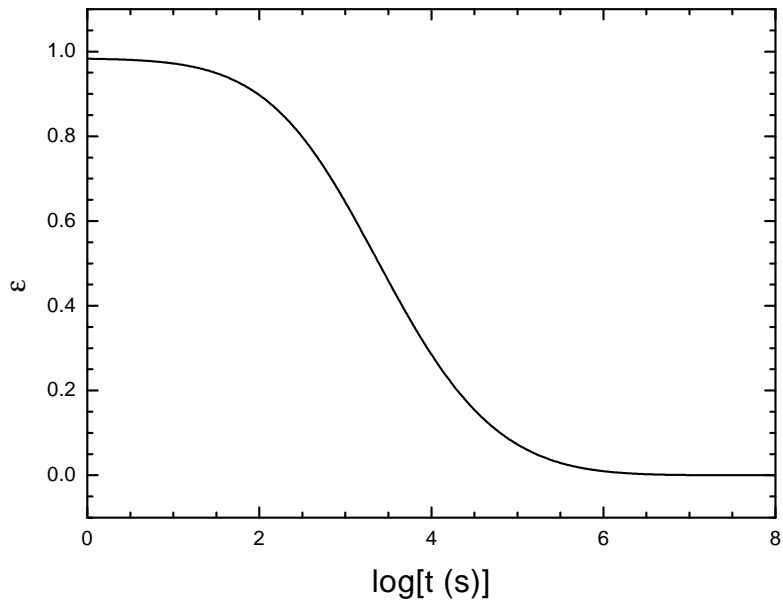}
   \caption{ Evolution of the radiative efficiency of the fireball ($\epsilon$).
   Parameters are the same as in Fig.~1.  }
   \label{}
   \end{center}
\end{figure}

\subsection{Numerical Results}

The evolution of the radius and swept-up mass are described by
(Huang et al. 1998a, 2000a, b)
\begin{equation}
dm = 4 \pi R^2 n m_{\rm p} dR,
\end{equation}
\begin{equation}
dR = \beta c \gamma (\gamma + \sqrt{\gamma^2 -1}) dt,
\end{equation}
where $t$ is the time measured in the observer's frame.
Then equations (\ref{improved generic model}) and (\ref{dudex})
can be solved numerically.

Figure 1 compares the evolution of the Lorentz factor calculated
according to equations (\ref{generic model}) and (\ref{improved
generic model}). In our calculations, we take $E_0 = 10^{52}$
ergs, $n = 1 $ cm$^{-3}$, $M_{\rm 0} = 2 \times 10^{-5}M_{\odot}$,
$\epsilon_e=1.0$, $\epsilon_B=0.01$. In both cases, equation
(\ref{epsilon}) is used to depict the evolution of $\epsilon$. We
see that, as expected above, the bulk Lorentz factor of the
fireball ($\gamma$) calculated by equation (\ref{improved generic
model}) (the solid line) declines more rapidly than that of
equation (\ref{generic model}) (the dashed line). But we notice
that the difference is slight. Figure 2 shows the time dependence
of  the  blastwave radius ($R$). Figure 3 shows the evolution of
the radiative efficiency of the realistic fireball ($\epsilon$).

The relation between the radius ($R$)
and the fireball momentum ($P=(\gamma^{2} - 1)^{1/2}$) is
shown in Figure 4. The solid line is the case when
$\epsilon$ evolves according to
equation (\ref{epsilon}). The dashed line is the adiabatic case,
i.e., $\epsilon \equiv 0$. The dotted line is the highly radiative
case, viz., $\epsilon \equiv 1$. We can see that, at early times
when the realistic fireball is ultra-relativistic
and highly radiative, the solid line
approximately satisfies $P \propto R^{-3}$.
At late times when the
fireball is non-relativistic and
adiabatic, the deceleration is approximately $P \propto R^{-3/2}$,
consistent with the Sedov limit.

\begin{figure}
   \epsscale{1.0}{1.0}          %% rescaled to {x_size}{y_size}, default is {1}{1}
   \plotone{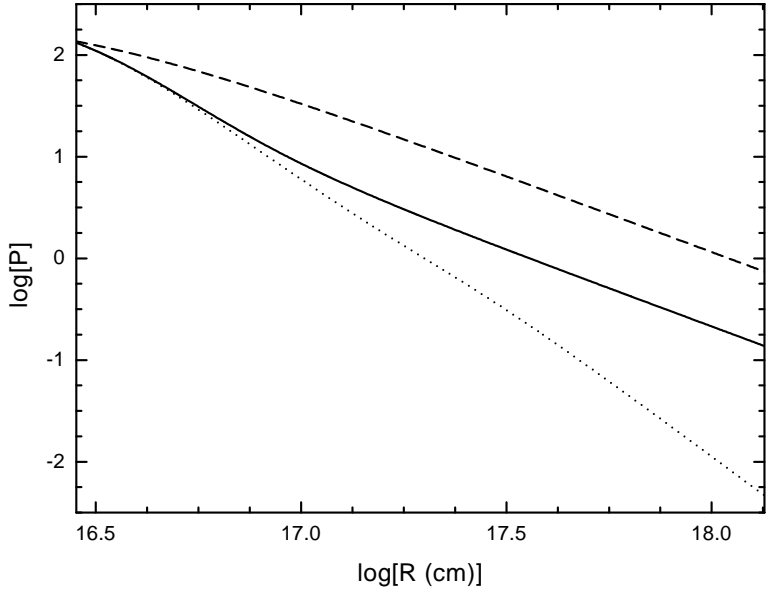}
   \caption{ Evolution of the fireball momentum ($P$) under different assumptions
   for the radiative efficiency of the fireball ($\epsilon$). The solid line is
   the case when $\epsilon$ evolves according to equation (\ref{epsilon}). The
   dashed line is the adiabatic case, i.e., $\epsilon \equiv 0$. The dotted line
   is the highly radiative case, viz., $\epsilon \equiv 1$. Parameters are the
   same as in Fig.~1. }
   \label{}
\end{figure}

We emphasize that for the $\epsilon \equiv const$ cases, the
results are precisely the same in the two models characterized by
equation (\ref{generic model}) and equation (\ref{improved generic
model}).

%%%%%%%%%%%%%%%%%%%%%%%%%%%%%%%%%%%%%%%%%%%%%%%%%%%%%%%%%%%%%%%%%
\section{light curve}

In section 2, the dynamical evolution of a postburst fireball has
been calculated numerically. As in Dai Huang \& Lu (1999), we
calculate the light curves of optical afterglows. The results are
shown in Figure 5. Here the solid line is drawn by using the
dynamics of equation (\ref{improved generic model}) and the dashed
line is drawn by using equation (\ref{generic model}). We see that
the difference between the two curves is not notable. Flux
densities on the dashed curve are higher by about 2 after the
light curve peak, but the slopes of the two curves are identical.
Please note that in our calculation, we have taken relatively
large parameter values for $\epsilon_e$ and $\epsilon_B$:
$\epsilon_e = 1.0, \epsilon_B = 0.01$. If these two parameters are
taken typical values as $\epsilon_e \sim 0.1, \epsilon_B \sim
10^{-4}$ --- $10^{-6}$, then the difference will be even smaller.

\begin{figure}
   \epsscale{1.0}{1.0}          %% rescaled to {x_size}{y_size}, default is {1}{1}
   \plotone{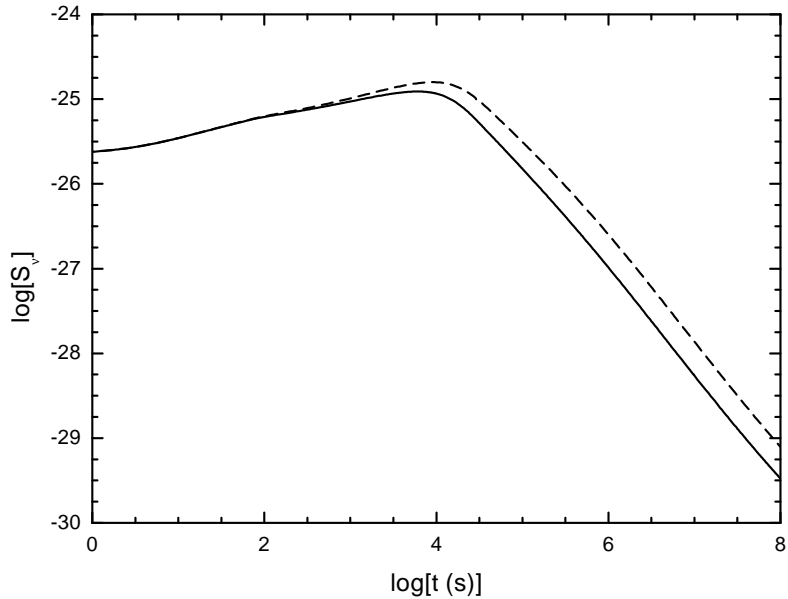}
   \caption{ Predicted afterglow light curves in fixed frequency $\nu = 10^{15}$Hz.
   $S_\nu $ is in units of ergs s$^{-1}$ cm$^{-2}$ Hz$^{-1}$. The dashed line
   corresponds to the generic model of HDL99, and the solid line corresponds
   to Eq.~(\ref{improved generic model}). Parameters adopted: $E_0 = 10^{52}$ ergs,
   $n = 1$ cm$^{-3}$, $M_{\rm 0} = 2 \times 10^{-5}M_{\odot}$, $\epsilon_e=1.0$,
   $\epsilon_B=0.01$, $p=2.1$, and $D=1$ Gpc. Note that after $\sim 10^{4}$ s,
   slopes of the two curves are approximately identical. }
   \label{}
\end{figure}

%%%%%%%%%%%%%%%%%%%%%%%%%%%%%%%%%%%%%%%%%%%%%%%%%%%%%%%%%%%%%%%%%
\section{Discussion and Conclusions}

The generic model of HDL99 is applicable to both radiative and
adiabatic fireballs, and in both ultra-relativistic and
non-relativistic phases. A possible problem is that whether this
model is correct or not when the radiative efficiency of the
blastwave ($\epsilon$) evolves with time. We have shown that in
this case, for the evolution of $\gamma$ and $R$, the errors
induced by the generic model is almost negligible. Errors in the
optical light curves are slightly amplified due to the strong
dependence of flux density on the Lorentz factor, but the results
are still acceptable. We suggest that the generic model in its
simple form of equation (7) in HDL99 could be safely used when
$\epsilon$ varies during the deceleration.

A dynamical model that is applicable to both relativistic
and non-relativistic expansion has been established for quasars
and active galactic nuclei by Blandford \& McKee (1977).
Their dynamics is most convenient for either adiabatic or
highly radiative blastwaves, even allowing for steady injection
of energy into the remnant from the central engine.
However, for partially radiative blastwaves,
especially blastwaves with an evolving efficiency, the simple
generic dynamical model of HDL99 is still more convenient.

%%%%%%%%%%%%%%%%%%%%%%%%%%%%%%%%%%%%%%%%%%%%%%%%%%%%%%%%%%%%
\begin{acknowledgements}

We thank an anonymous referee for helpful comments and suggestions.
This work was supported by The Foundation for the Author of
National Excellent Doctoral Dissertation of P. R. China (Project
No: 200125), the Special Funds for Major State Basic Research
Projects, the National Natural Science Foundation of China, and
the National 973 Project (NKBRSF G19990754).

\end{acknowledgements}

\end{document}